\documentclass[prl,onecolumn,amsmath,amssymb,superscriptaddress,floatfix]{revtex4}
\usepackage{graphicx,epsf}
\usepackage{bm}

\begin{document}

\markboth{Ana L. C. Pereira and Peter A. Schulz}
{Graphene in the Quantum Hall Regime:  Effects of Vacancies, Sublattice Polarization and Disorder}

\vspace{2cm}
\title{Graphene in the Quantum Hall Regime: \\ Effects of Vacancies, Sublattice Polarization and Disorder}
\vspace{2cm}
\author{Ana L. C. Pereira}
\author{ Peter A. Schulz}

\address{Instituto de F\'{\i}sica, Universidade Estadual de Campinas - UNICAMP -
C.P. 6165, 13083-970, Brazil
\\  analuiza@ifi.unicamp.br }

%\begin{history}
%\received{Day Month Year}
%\revised{Day Month Year}
%\accepted{(Day Month Year)}
%\comby{(xxxxxxxxxx)}
%\end{history}

\begin{abstract}
We investigate the effects of vacancies, disorder and sublattice polarization on the electronic properties of a monolayer graphene in the quantum Hall regime.  Energy spectra as a function of magnetic field and the localization properties of the states within the graphene Landau levels (LLs) are calculated through a tight-binding model.  We first discuss our results considering vacancies in the lattice, where we show that vacancies introduce extra levels (or well-defined bands) between consecutive LLs. An striking consequence here is that extra Hall resistance plateaus are expected to emerge when an organized vacancy superlattice is considered.
Secondly, we discuss the anomalous localization properties we have found for the lowest LL, where an increasing disorder is shown to enhance the wave functions delocalization (instead of inducing localization). This unexpected effect is shown to be directly related to the way disorder increasingly destroys the sublattice (valley) polarization of the states in the lowest LL. The reason why this anomalous disorder effect occurs only for the zero-energy LL is that, in absence of disorder, only for this level all the states are sublattice polarized, i.e., their wave functions have amplitudes in only one of the sublattices.
\end{abstract}


\maketitle

\section{1. Introduction}

The quantum Hall effect (QHE) observed in graphene few years ago\cite{novoselov,zhang1} presents significant differences compared to the QHE in usual two-dimensional electron gases. Due to the unique features of the Dirac-like band structure of graphene around Fermi energy, the sequence of plateaus is shifted by half-integer if compared to the usual QHE, and the energy dependence of the Landau levels (LLs) with magnetic field ($B$) is not linear anymore, but goes with root square of ($B$). Another important difference is that, in addition to the spin degree of freedom, graphene LLs exhibit a valley degree of freedom, which has been the subject of many interesting studies and application proposals\cite{zhang2,lederer,levitov,ando,altshuler,sheng2,chak,beenakker2}.

In this paper we discuss the effects of considering disorder and vacancies on the electronic properties of graphene in the presence of $B$, reviewing and expanding the discussions of recent results on these subjects \cite{ana_valley,ana_vac}.
We calculate energy spectra as a function of magnetic field, density of states (DOS) and localization properties of the states within the graphene LLs. The tight-binding lattice model used is described in Section 2. In Section 3 we analyse the role of vacancies in the lattice, focusing on the evolution with $B$ of the localized states introduced in the DOS by the vacancies, and on the interesting consequences of considering an increasing density of vacancies\cite{ana_vac}. In Section 4, we discuss the relation we found between anomalous localization properties in the zero-energy LL and the way that disorder destroys the sublattice polarization presented by the states within this LL in absence of disorder\cite{ana_valley}.

\section{2. Model}

%%%%%%%%%%%%%%%%%%%%%%% Figure 1 %%%%%%%%%%%%%%%%%%%%%%%

\begin{figure}[t]

\vspace{1cm}
\includegraphics[width=16cm]{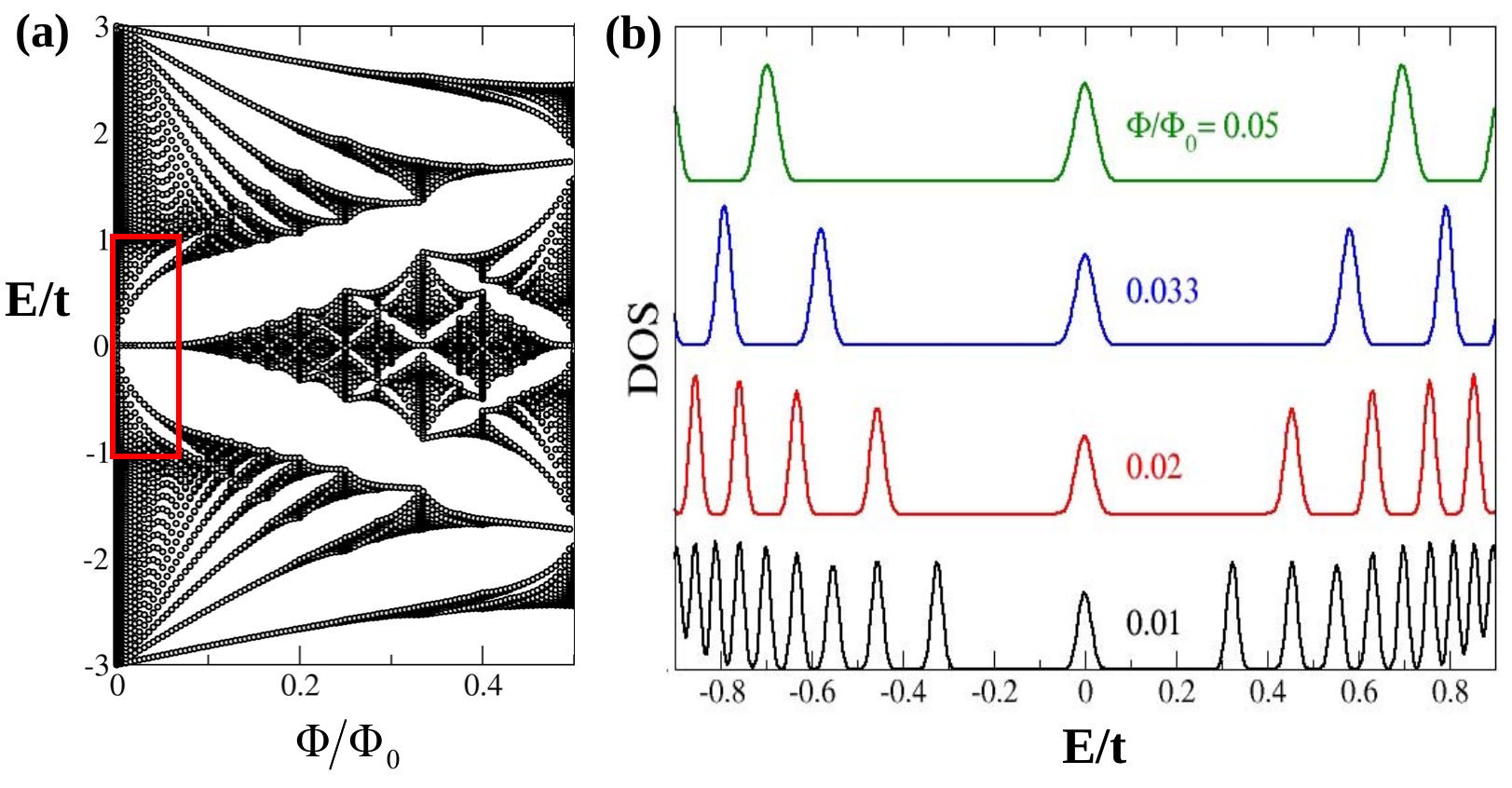} \hspace{-0.5cm}

%\vspace{-1.6cm}
\caption{{\bf (a)} Energy spectrum as a function of magnetic flux for the hexagonal lattice. Rectangular region identifies the continuum limit of the spectrum, corresponding to the LLs of a monolayer graphene. {\bf (b)}  Density of states for four different values of magnetic flux, shifted vertically for clarity. The LLs  are broadened by a white-noise disorder with $W/t$=0.1. }
\end{figure}

%%%%%%%%%%%%%%%%%%%%%%%%%%%%%%%%%%%%%%%%%%%%%%%%%%%%%%%%%

The numerical approach used in this work to calculate the electronic properties of graphene monolayers consists of a tight-binding model with nearest neighbors hoppings in a honeycomb lattice.
The non-interacting Hamiltonian model is the following:

\begin{equation}
H = \sum_{i} \varepsilon_{i} c_{i}^{\dagger} c_{i}
+ t  \sum_{<i,j>} (e^{i\phi_{ij}} c_{i}^{\dagger} c_{j} + e^{-i\phi_{ij}}
c_{j}^{\dagger} c_{i})
\end{equation}

\hspace{-\parindent}where $c_{i}$ is the fermionic operator on site $i$.
A perpendicular magnetic field $B$ is included by means of a Peierls' substitution in hopping parameter ($t$$\approx$2.7eV for graphene): $\phi_{ij}= 2\pi(e/h) \int_{j}^{i} \mathbf{A} \! \cdot \! d \mathbf{l} \;$. In the Landau gauge, $\phi_{ij}\!=\!0$ along the $x$ direction and $\phi_{ij}\!=\pm \pi (x/a) \Phi / \Phi_{0}$
along the $\mp y$ direction, with $\Phi / \Phi_{0}=Ba^{2}\sqrt{3}e/(2h)$  ($a$=2.46{\AA} is the lattice constant for graphene).

Square or rectangular graphene lattices are considered, of up to 8.000 atomic sites, which are repeated by means of periodic boundary conditions. The electronic spectrum as a function of magnetic flux obtained for the honeycomb lattice is depicted in Fig. 1(a). We focus on the low-flux limit with energy window around the Dirac point (identified by the rectangular area), where the LLs are well defined.

For the sake of Landau bands broadening in the DOS calculations, as shown in Fig. 1(b), we consider disorder through on-site white-noise fluctuations, by sorting uncorrelated orbital energies within $\varepsilon_{i} \leq |W/2|$.
When simulating vacancies, the hopping parameters around the vacancy are set to zero and the on-site energy at the vacancy site is made equal to a large value outside the energy range of the DOS. Atomic relaxation and rebounding around the monovacancies simulated are not considered, however, these effects are not expected to modify qualitatively the results we discuss\cite{ana_vac}.

The localization properties of the states within the LLs are inspected by means of  the participation ratio (PR)\cite{thouless},

\begin{equation}
PR = \frac{1}{N \sum_{i=1}^{N}|\psi_{i}|^{4}},
\end{equation}

\hspace{-\parindent}where $\psi_{i}$ is the amplitude of the normalized wave function on site $i$, and $N$ is the total number of lattice sites. The PR gives the proportion of $N$ sites over which the wave function is spread. In this way, the PR for a localized state will tend to zero in the thermodynamic limit.

\section{3. Vacancy Superlattices and New Quantum Hall Plateaus}

We consider vacancies in graphene lattices in the presence of a perpendicular magnetic field. Previous to our work\cite{ana_vac}, the effects of vacancies on graphene have been studied mainly in situations in the absence of magnetic field\cite{castroneto1,castroneto2008,brey2} and results on the graphene electronic structure modifications due to vacancies in the quantum Hall regime were sparse\cite{castroneto2}.

We show in Fig. 2 that vacancies give origin to new states (open circles) which appear between consecutive the LLs in the energy spectra. The energies of these vacancy states are increased with increasing magnetic fields. If two single vacancies are introduced in the lattice far away from each other, two degenerated states are introduced between LLs at the same energy positions shown in Fig. 2(b) for a single vacancy. However, a different scenario emerges if the two vacancies are introduced close to each other as revealed by the energy splitting observed on the energy-magnetic-flux spectra shown in Figs. 2(c) and 2(d), for two different inter-vacancies distances: $D$$=$11$a_{cc}$ and $D$$=$5$a_{cc}$, where $a_{cc}$=$1.42$\AA $\,$ is the distance between two neighbor carbons in the lattice. One can see that this energy splitting increases with decreasing vacancy distances or decreasing magnetic field. These results show that sufficiently near vacancies lead to the coupling of
 the states localized around each vacancy.

%%%%%%%%%%%%%%%%%%%%%%% Figure 2 %%%%%%%%%%%%%%%%%%%%%%%

\begin{figure}[t]
\vspace{0.1cm}
\includegraphics[width=15.6cm]{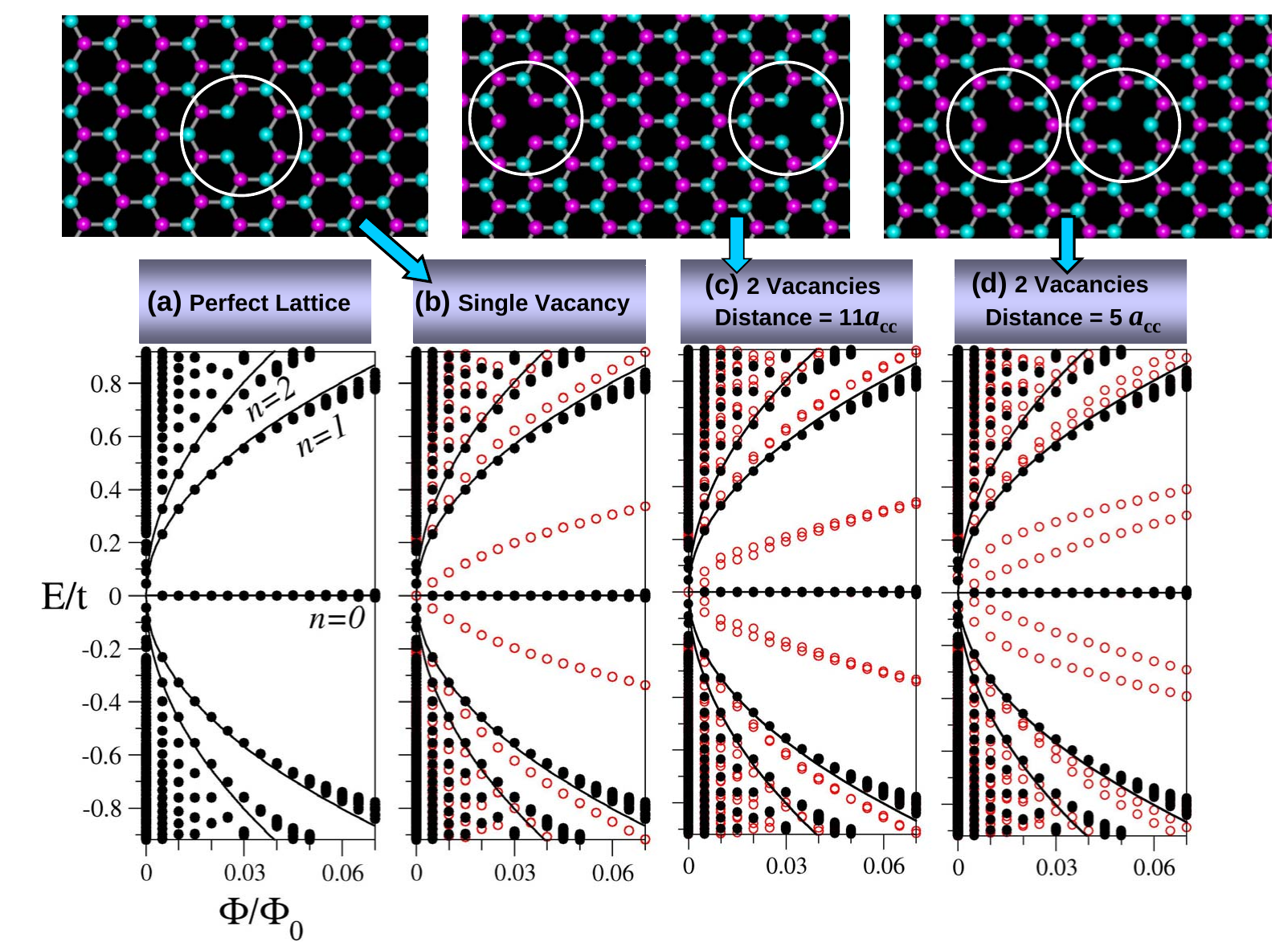}
\vspace{0.3cm}
\caption{ Lattice representation and the corresponding energy-magnetic flux spectra for {\bf (a)} a perfect graphene lattice, {\bf (b)} lattice with a single-vacancy, {\bf (c)} lattice with two vacancies, separated by a distance of 11 $a_{cc}$, and {\bf (d)} lattice with two vacancies, separated by a distance of 5 $a_{cc}$, where $a_{cc}=1.42$\AA $\,$ is the distance between two neighbor carbon atoms.  Continuum lines show the energy dependencies with $\sqrt{\Phi/\Phi_0}$  of the $n$=0,$\pm$1,$\pm$2 graphene LLs, for guidance. Vacancies in the lattice introduce states with energies between the LLs, which are shown as open circles.}
\end{figure}

%%%%%%%%%%%%%%%%%%%%%%%%%%%%%%%%%%%%%%%%%%%%%%%%%%%%%%%%%

We also show that sufficiently high density of vacancies (organized vacancy superlattice) introduces an extra band in between LLs with a non-negligible DOS (Fig. 3(a)). Vacancies are natural defects or can also be formed after ion-beam irradiation: in this sense, densities of vacancies are expected to be found in a random distribution pattern. However, due to the advantageous electronic properties of graphene\cite{geim_rev}, microstructuring efforts have experienced a fast recent development. Having in mind a promising scenario of microscopic manipulation (attempts of tailoring graphene with scanning probe microscope lithography have recently been reported\cite{biro,maan}), an ordered vacancy network, introduced on purpose by following a previous design, is worth to be considered.

%%%%%%%%%%%%%%%%%%% Figure 3 %%%%%%%%%%%%%%%%%%%%%%%

\begin{figure}
\begin{center}
\vspace{-3cm}
\includegraphics[width=18cm]{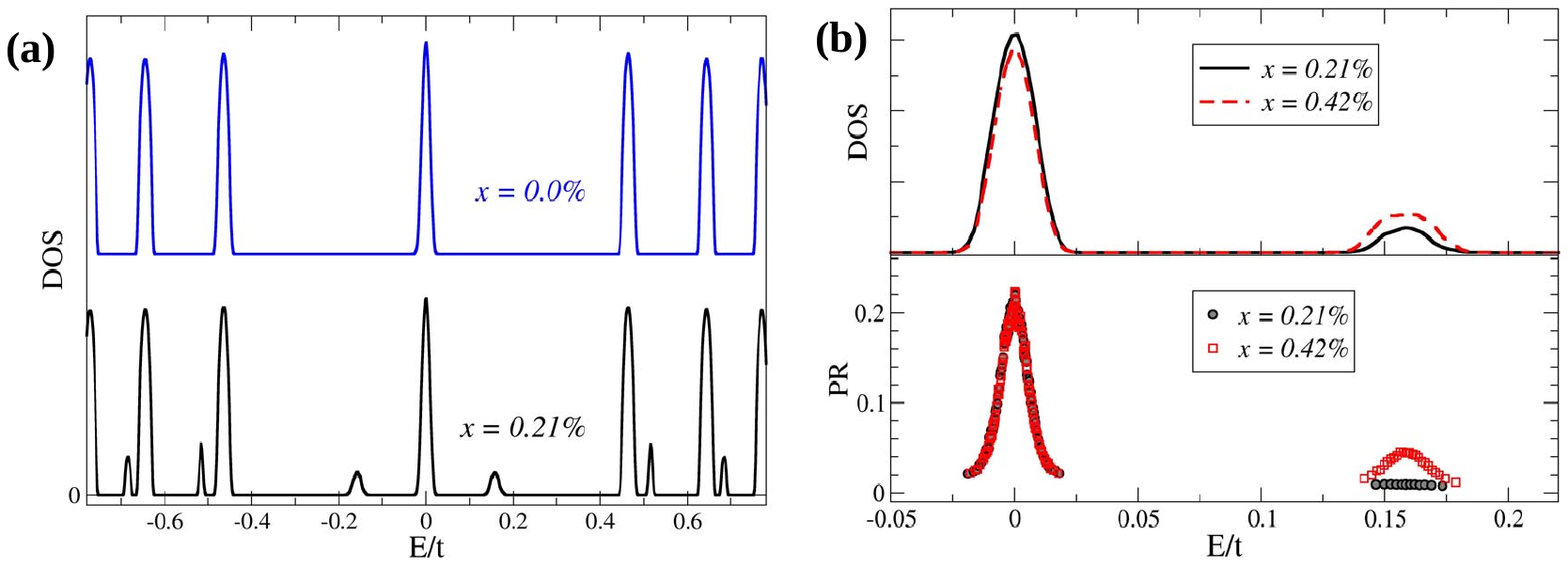}
\vspace{-1.5cm}
\end{center}
\caption{ {\bf (a)} DOS for a graphene lattice without vacancies compared with a lattice containing a rectangular arrangement of 12 vacancies (vacancy-superlattice), corresponding to a vacancy concentration of  $x$=$0.21\%$, shifted vertically for clarity.
The formation of extra bands between LLs, due to the vacancies, is evident. LLs are broadened by a white-noise disorder with $W/t$=0.1 and $\Phi/\Phi_0=2/96\approx 0.02$. {\bf (b)} DOS and PR for states within the zero-energy LL and the first vacancy-related band. Increasing vacancy concentration ($x$) leads to a modulation of the PR within the extra band similar to that of the LL. [Figure reproduced from Ref. 12].}
\end{figure}

%%%%%%%%%%%%%%%%%%%%%%%%%%%%%%%%%%%%%%%%%%%%%%%%%%%%%

The most interesting feature is that the localization properties of these extra bands are similar to that of the LLs, which could lead to new magneto-transport features, such as extra Hall resistance plateaus.
To inspect the localization properties of the states within the extra vacancy-bands, we use the participation ratio (PR), defined in Eq. (2).
In Fig. 3(b), we focus our attention only on the zero-energy Landau band ($n$=$0$) and on the vacancy-band between the $n$=$0$ and $n$=$1$ LLs. From the PR calculation, one can see that if the concentration of vacancies is low ($x=0.21\%$), all the states in the vacancy band are localized (the PR for a localized state tends to zero). On the other hand, if the intervacancy distances diminish,  these vacancy states may be effectively coupled and eventually become delocalized. Strikingly, the PR for a vacancy concentration of $x=0.42\%$ shows a strong modulation for the vacancy related band, similar to the ones for the Landau bands. This behavior signalizes the existence of new delocalized states in the DOS. In this way, a vacancy lattice could be identified in magnetotransport measurement by extra quantum Hall plateaus in between the original plateaus.

The vacancy concentration of $x$=$0.42\%$ imposes a vacancy lattice with distances between vacancies of $D \approx 18 a_{cc}$ in both lattice directions. For both vacancy concentrations shown in Fig. 3, all the vacancies were introduced in the same sublattice. However, we emphasize that we also tested configurations (not shown here) where the vacancies are equally distributed between the two sublattices (zero degree of uncompensation), and exactly the same effect seen in Fig. 3 is observed, i.e., after a critical concentration, the PR clearly indicates a delocalization of the states in the middle of the vacancy-band.

%%%%%%%%%%%%%%%%%%%%%%%%%%%%%%%%%%%%%%%%%%%%%%%%%%%%%%%%%%%%%%%%%%%%%%%%%%%%%%
%%%%%%%%%%%%%%%%%%%%%%%%%%%%%%%%%%%%%%%%%%%%%%%%%%%%%%%%%%%%%%%%%%%%%%%%%%%%%%

%\vspace{3.5cm}

\section{4. Anomalous Localization and Sublattice Polarization}

Including disorder in the graphene lattice through random diagonal disorder models (white-noise and Gaussian-correlated disorder), we observed that anomalous localization properties occur in the zero-energy LL, where increasing disorder enhance the participation ratio of the states, instead of inducing the expected wave function localization\cite{ana_valley}. We found that this effect is related to the fact that in the absence of disorder (ideal graphene), all the states within the lowest LL are sublattice polarized, i.e., their wave functions have non-zero amplitudes in only one of the sublattices. Here we expand the discussion from our previous work\cite{ana_valley}, showing in more details how the inclusion of disorder destroys the sublattice polarization in the $n$=0 LL.

Figure 4 summarizes our main results, showing the evolution of the DOS,  the participation ratio (PR) and the sublattice polarization around the Dirac point, for increasing disorder in the white-noise disorder model\footnote{While here we illustrate the anomalous localization effect showing only results for the white-noise disorder model, in Ref. 11 a comparison with a Gaussian-correlated disorder model is carefully discussed.}.
From the DOS, one can see the lowest three LLs ($n$=0,1,2) and their increasing broadening with the disorder.
These results have been calculated on unit cells of 60$\times$60 sites (60 zig-zag chains, each containing 60 sites), with periodic boundary conditions, considering  $\Phi / \Phi_0$=$1/30$. Averages over hundred disorder realizations were undertaken.
Extended states are identified by peaks in the PR around the center of each LL\cite{pereira}.
An important effect we want to point out in Fig. 4 is the anomalous increase of PR with increasing disorder, occurring for states of the $n$=0 LL. One can easily observe that the maximum values of PR corresponding to the $n$=0 LL are substantially increased for disorder amplitudes from $W/t$=0.05 to 0.5. This behavior is surprising and called anomalous because increasing disorder is expected to have the opposite effect: making states more localized.

%%%%%%%%%%%%%%%%%%%%%%% Figure 4 %%%%%%%%%%%%%%%%%%%%%%%

\begin{figure}[t]
\vspace{0.7cm}
\includegraphics[width=10cm]{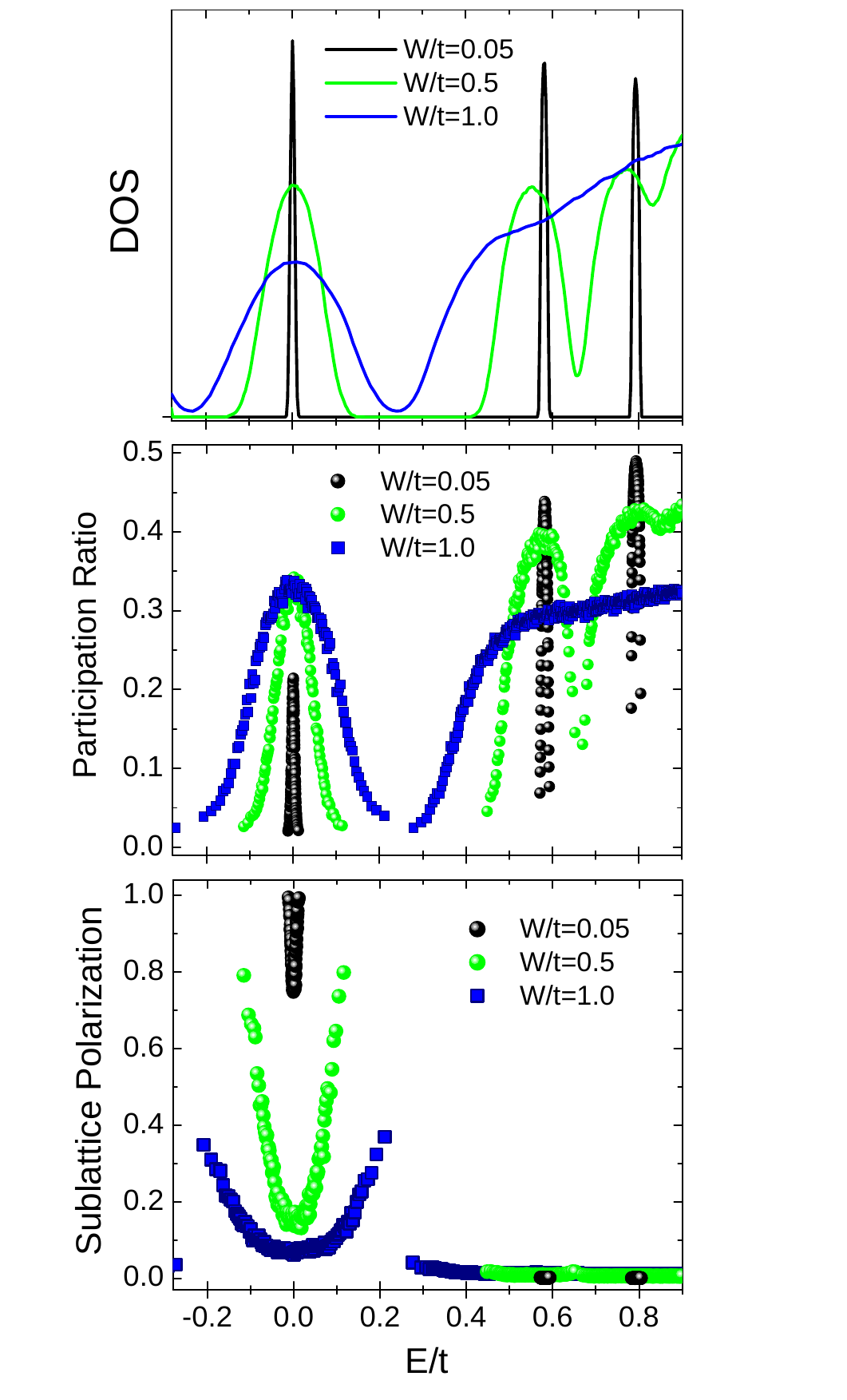}
%\vspace*{8pt}
\caption{DOS for three different disorder amplitudes (broadening of LLs increases with
disorder), and the corresponding participation ratio, and sublattice polarization.
Three LLs are shown ($n$=0,1,2) for flux $\Phi / \Phi_0$=$1/30$. [Figure reproduced, in part, from Ref. 11].}
\end{figure}

%%%%%%%%%%%%%%%%%%%%%%%%%%%%%%%%%%%%%%%%%%%%%%%%%%%%%

To inspect the uncompensation on the distribution of the wave functions amplitudes between the two sublattices, we define the sublattice polarization as:

\begin{equation}
Polarization = ||\Psi_{A}|^2-|\Psi_{B}|^2|,
\end{equation}

\hspace{-\parindent}where $|\Psi_{A(B)}|^2$ is the sum of wave function amplitudes only over sites of sublattice A(B). Wave function normalization implies $|\Psi_A|^2 +|\Psi_B|^2 = 1$.  It is important to notice that this defined sublattice polarization gives the difference in modulus of the contribution from each sublattice to the total probability density of the state. In this way, Polarization=0 in Fig 4 means that the state is equally distributed on both sublattices, while Polarization=1.0 means that the state has amplitudes over only one of the sublattices.
In Fig.4, one can see that the states from the $n$=0 band become increasingly unpolarized with disorder, and clearly show an interesting ``{\it U-shape}" polarization profile: states at the band tails - strongly localized (as observed from the PR) - remain more polarized than the delocalized ones, from LL center.

%%%%%%%%%%%%%%%%%%%%%%% Figure 5 %%%%%%%%%%%%%%%%%%%%%%%

\begin{figure}[bt]
\vspace{0.5cm}
\includegraphics[width=14cm]{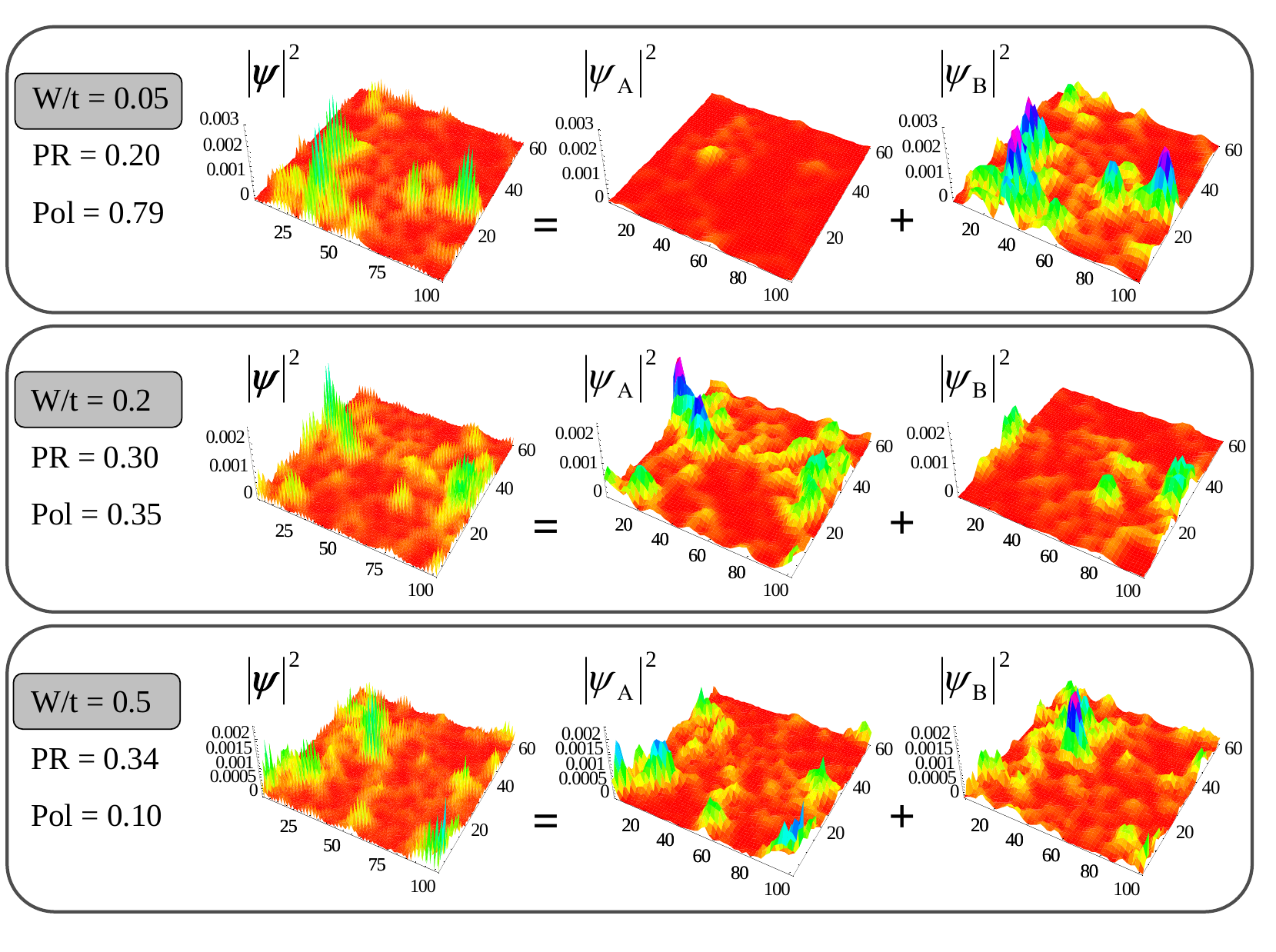}
\vspace{0.5cm}
\caption{Wave functions from the $n$=0 LL, and their decomposition into the two subattices, for three different values of disorder. Wave function amplitudes  ($|\Psi|^2$) are firstly shown over all the 60$\times$102 lattice sites, and then contributions from each
sublattice ($|\Psi_A|^2$ and $|\Psi_B|^2$) are plotted separately. Disorder increasingly promotes valley mixing, which causes the observed decreasing of sublattice polarization (Pol) for the states of this LL. Results are for $\Phi / \Phi_0$=$1/30$.}
\end{figure}

%%%%%%%%%%%%%%%%%%%%%%%%%%%%%%%%%%%%%%%%%%%%%%%%%%%%%

In Figure 5 we show three examples of wave functions (probability densities), $|\Psi|^2$, and their decompositions into the two sublattices, $|\Psi_A|^2$ and $|\Psi_B|^2$. The examples correspond to typical extended states from the middle of the $n$=0 LL, for three different intensities of disorder. PR and sublattice polarization (Pol) values for each state are indicated on the figure.
One can see that for the smallest disorder shown in Fig. 5 ($W/t$=0.05) the amplitudes of the wave function of the chosen state over the sublattice A are much less significant than over sublattice B (note that the amplitude scales are the same in the decomposition), and in fact, we calculate a 79$\%$ sublattice polarization for this state. As disorder is increased, we see that sublattice-polarized wave functions from the n=0 LL increasingly spread to both sublattices, allowing in this way more sites to participate in the wave function. This naturally produces an easier percolation for the wave function across the entire system, explaining the anomalous increase in the PR of the state with increasing disorder.

%%%%%%%%%%%%%%%%%%%%%%% Figure 6 %%%%%%%%%%%%%%%%%%%%%%%

\begin{figure}[bt]
\includegraphics[width=13cm]{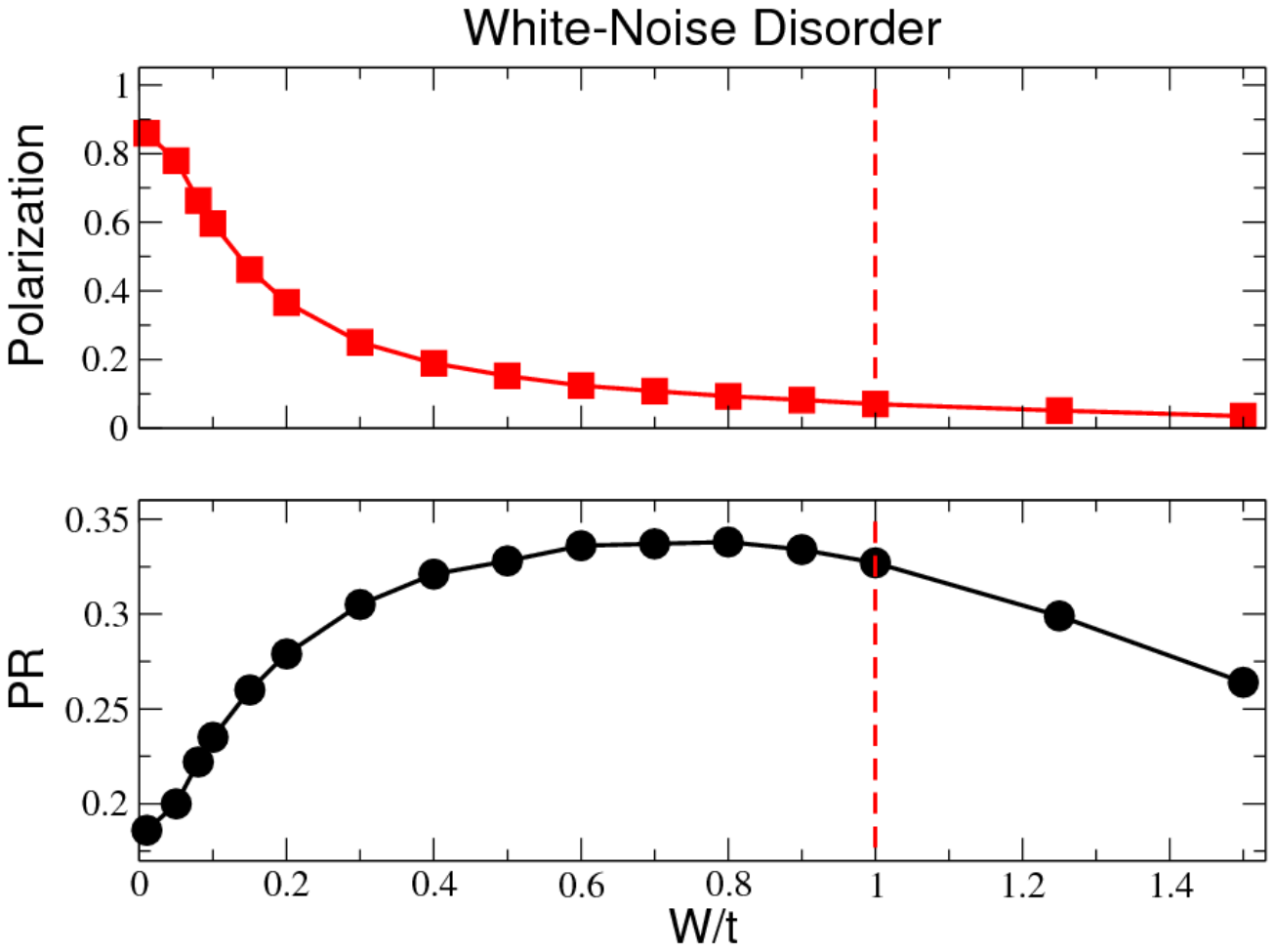}
\vspace*{8pt}
\caption{Values of valley polarization and participation ratio (PR) for the states at Dirac point (at the center of the $n$=0 LL), as a function of white-noise disorder amplitude. Dashed line indicates the $W/t$ for which the broadening of LLs is such that the $n$=0 LL starts to overlap with neighbors bands (the band tails start to touch each other). [Figure reproduced from Ref. 11].}
\end{figure}

%%%%%%%%%%%%%%%%%%%%%%%%%%%%%%%%%%%%%%%%%%%%%%%%%%%%%

Fig. 6 allows a more detailed observation of the clear correspondence between the anomalous increase of the PR with disorder and the gradual extinguishment of sublattice polarization due to disorder. PR and polarization values at the middle of the $n$=0 LL, i.e., at E=0, are shown as a function of disorder, and we see that the increase in the PR is accompanied exactly by the same rate of decrease in the polarization.
One can also see from Fig. 6 that  the expected and well-known localization with increasing disorder only comes back to this scenario after disorder has significantly mixed valleys.

\section{5. Conclusions}

We showed that vacancies in graphene lattices introduce new states in the DOS, with energies between Landau levels. These states can be assessed by usual switching of the magnetic field or gate voltages in quantum Hall measurements, but are not expected to reveal any fingerprint in these measurements because of their strong localized character. Sufficiently near vacancies lead to the coupling of the states localized around each vacancy. We found that an organized vacancy superlattice introduces an extra band in between LLs, with a non-negligible DOS. The localization properties of these extra bands are similar to that of the LLs, which could lead to new magneto-transport features, such as extra Hall resistance plateaus.

We also discussed the anomalous localization properties for the states from the zero-energy LL.
The evolution of the degree of localization (participation ratio) with disorder were followed simultaneously with the sublattice polarization of these states. We showed in this way, that the role of disorder in destroying sublattice polarization in the $n$=0 LL reveals to be directly connected to the anomalous increase of PR with disorder.

\vspace{0.5cm}

\section{Acknowledgements}

This work is supported by FAPESP and CNPq.

%\vspace{1cm}

\end{document}